\documentclass[10.8pt,a4paper,twoside]{article}
\usepackage{float}
\usepackage{multirow}
\usepackage{pdflscape}
\usepackage{caption,subcaption}
\usepackage[final]{graphicx}
\usepackage{lmodern} 
\usepackage{graphicx} 
\usepackage{amssymb,amsthm,amsmath, amsfonts}

\graphicspath{{images/}}

\usepackage[noadjust]{cite}
\usepackage{filecontents}

\usepackage{color}
\usepackage{lscape}
\usepackage{setspace}

\begin{document}

\baselineskip=0.050in
{\bf \Large
\begin{center}
Analytical comparison between $X(3)$ and $X(5)$ models of the Bohr Hamiltonian
\end{center}

}
\begin{center}
{\bf K.R. Ajulo$^{*+}$}\footnote{$^{*}$E-Mail:~ 19-68eo001@students.unilorin.edu.ng}, {\bf K.J. Oyewumi$^{+}$}\footnote{E-Mail:~ kjoyewumi66@unilorin.edu.ng}

{$^{1,2}$Faculty of Physical Science, University of Ilorin, P.M.B, 1515, Kwara State, Ilorin, Nigeria.}
\end{center}

\begin{abstract}
\noindent
The 3-D Bohr-Mottelson Hamiltonian for $\gamma$-rigid prolate isotopes, known as $X(3)$, is solved via inverse square potential having only one free parameter, $\beta_{0}$. The exact form of the wave functions and the energy spectra are obtained as a function of the free parameter of the potential that determines the changes in the spectra ratios and the $B(E2)$. Since $X(3)$ is an exactly separable $\gamma$-rigid version of $X(5)$, the solutions are compared with the $X(5)$ model and some new set of equations that show the relationships between the two models are stated. In other to show the dynamical symmetry nature of the solutions, the entire solutions from $\beta_{0}=0$ to $\beta_{0}=\infty$ are compared with $U(5)$, $X(5)$ and $SU(3)$. The solutions spread from the region around $U(5)$ over $X(5)$ and approach $SU(3)$ at $\beta_{0}=\infty$. The exact solutions obtained via variational procedure are compared favourably with some existing $X(3)$ models found in the literature.  The strong agreement between the present model and $X(3)$ via infinite square well potential is discussed.  Twelve best critical point  isotopes, $^{102}$Mo, $^{104-108}$Ru,  $^{120-126}$Xe,  $^{148}$Nd, $^{184-188}$Pt are chosen for experimental realization of the model and moderate agreements are recorded. An excellent agreement which appears in the first $\beta$-excited state in the comparison of the present model with three  $N=90$ isotones: $^{150}$Nd, $^{154}$Gd, and $^{156}$Dy, known to be  $X(5)$ candidates,  suggests that the present model compensates the $X(5)$ models whose  predictions are excellent in the ground states but moderately bad in the first $\beta$-excited states.  
 
\end{abstract}


\section{ Introduction}
The establishment of the critical point symmetries by Iachello$^{1,2}$  paved way for the shape phase changes between various dynamical symmetries. Somehow, these critical point symmetries$^{1,2}$ are not pure dynamical symmetries because of the group reduction stated in  the interacting boson model (IBM) structure of Iachello and Arima$^{3}$. However, compensating for this, fitting procedure  given by nearly the same shapes having the same potential description in the geometrical collective model$^{4,5}$ is required. Two types of these critical symmetries are known: one, the $U(5)\rightarrow O(6)$ dynamical symmetry that describes the phase changes from the spherical vibrator shape phase$^{6}$ to the  $\gamma$-unstable nuclei$^{7}$; two, the $U(5)\rightarrow SU(3)$ which is the  phase changes from the spherical vibrator shape phase$^{6}$ to the dynamical symmetry of axial rotors. The two resulted models are respectively $E(5)$  and $X(5)$, both stated in ref.$^{2}$. The $E(5)$ coincides with the phase changes of the second-order shape phase while the $X(5)$ coincides with the phase changes of the first-order shape phase.

\noindent
$X(3)$ model which was first proposed by Bonatsos$^{8}$ and recently presented by others$^{9-12}$ which  is  an exactly separable  $\gamma$-rigid form of the $X(5)$ critical point symmetry$^{2}$. One of the physical significance of  the $X(3)$ model is that it reveals the resemblance in the  $\beta$-bands between the $X(5)$ and the $E(5)$ predictions.  It is achieved by the exact separation of angular variables and  shape. The $X(3)$ model is defined  by the collective coordinate $\beta$ and two Euler angles since  $\gamma$ is assumed to be zero, unlike in the case of $X(5)$, where $\gamma$ is varied around  $\gamma^{0}=0$  in the harmonic oscillator potential$^{1}$. This implies that, only three variables: $\beta$ and $\theta_{i}$ are involved in the $X(3)$ model. Effortlessly, the present work treats  nuclei  as $\gamma$-rigid, with the axially symmetric prolate shape obtained at  $\gamma^{0}=0$ and easily achieves an exact separation of the $\beta$ variable from the Euler angles. This exact separation of the shape variables can not be found in  other shape phase changes and also in $X(5)$.  

\noindent
Recently, in the literature$^{8-12}$, the Bohr-Mottelson Hamiltonian for $X(3)$ model have been constructed, solved using different potential domains and has been referred to as $\gamma$-rigid version of $X(5)$. However, the analytical comparison between the wave functions, the energy spectra, critical orders, spectra ratios and the electromagnetic transitional probabilities, $B(E2)$, in the ground states and in the $\beta$-excited states have not yet been presented. Albeit, in the recent work$^{13}$, the $X(3)$ solutions have been compared with the $X(5)-{\beta^{2}}$, $X(5)-{\beta^{4}}$ in the framework of interacting bosom model (IBM) and also compared with the experimental data, the  analytical comparisons between the $X(5)$ and $X(3)$  have not been shown. 

\noindent
Analytically, the present work presents the similarities and the differences between the $X(5)$ and its $\gamma$-rigid version, $X(3)$ model. It employs the importance of a one-sided bound and one parameter inverse square potential$^{14}$  to provide the  domains for the  Bohr-Mottelson Hamiltonian solutions. The potential has a linear slope wall just like some nuclear potentials such as the Davidson potential$^{15}$ used in ref.$^{16}$, the Kratzer potential$^{17}$ employed in ref.$^{18}$, the Morse potential used in refs$^{19,20}$ and others. However, the solvability  of the potential in this model is such that, the differential equation containing  $\beta$ is analytically solvable since  the restriction of $\gamma$ to zero causes the kinetic energy term of the Hamiltonian to reduce certain number of variables and  the usual  Bohr-Mottelson Hamiltonian$^{21,22}$ becomes the Davydov-Chaban Hamiltonian$^{23}$ where the  elemental volume is now proportional to $\beta^{3}$. Consequently, the vibrational kinetic energy operator of the Hamiltonian, 

\begin{equation}
\hat{H} =\hat{T}_{tot} +V,
 \end{equation}
reads$^{9}$
\begin{equation}
 \hat{T}_{vib} = -\dfrac{\hbar^{2}}{2B}\dfrac{1}{\beta^{2}} \dfrac{\partial}{\partial \beta} \beta^{2} \dfrac{\partial}{\partial \beta}
 \end{equation} 
and the rotational kinetic energy operator$^{9}$ reads
\begin{equation}
 \hat{T}_{rot} = \dfrac{\hat{Q}^{2}}{6B \beta^{2}}, 
 \end{equation} 
where $\hat{T}_{tot}$ is the sum of the vibrational   and the rotational kinetic energy operators, $\hat{Q}^{2}$ represents the angular momentum in the intrinsic frame of reference and $B$ is the mass parameter. The inverse $\beta$ square of the potential can be easily absorbed by the $\hat{T}_{rot}$ term leaving out the single free potential parameter, $\beta_{0}$, to play out in the phase shape transition. The two conditions attached to the potential are very important in the description of the energies  and the phase shape transition of the nuclei. The one-parameter inverse square potential$^{14}$ chosen is of the form
\begin{equation}
V(\beta)=\begin{cases}
\dfrac{\beta_{0}}{\beta^{2}}, \text{ if } 0\leq \beta \leq \beta_{0},\\
\infty, \text{ if } \beta > \beta_{0},
\end{cases}
\end{equation}
where the free parameter, $\beta_{0}$, is also referred to as the variation parameter. It variation affects the shape phase transitions and the signatures of the nuclei. In a more simpler explanation, it is expected that the solutions should shift forward as the $\beta_{0}$ shifts forward and solutions should shift backward as the $\beta_{0}$ shifts backward.
 A typical inverse square potential is bound on the left and unbound on the right, and it has a minimum at some positive values of $\beta_{0}$ that forces the particles to infinity as $\beta_{0}\rightarrow 0$. As a result, the particle's energy states is  one-sided, with energies escaping through the unbound side. 
\noindent
Therefore, with the potential in Eq.(4), the Davydov-Chaban Hamiltonian$^{23}$ connected to the prolate rigid nuclei can be written as

\begin{equation}
\hat{H} = -\dfrac{\hbar^{2}}{2B} \dfrac{1}{\beta^{2}} \dfrac{\partial}{\partial \beta} \beta^{2} \dfrac{\partial}{\partial \beta} -\dfrac{\hbar^{2}}{2B}\left(\dfrac{\beta_{0}}{\beta^{2}}- \dfrac{\hat{Q}^{2}}{3\hbar^{2} \beta^{2}}\right).  
\end{equation}
Now it has been stated  that the Bohr-Mottelson Hamiltonian explains the collective motion of the nuclei, the roles of   the potential energy and its associated parameter, $\beta_{0}$,  in the description of phase shape transitions will be seen later in the result. Although this is not the aim of this paper, the roles $\beta_{0}$ plays even as we compare $X(3)$ and $X(5)$ can not be left out without being discussed.

\section{ Methodology of $X(3)$ model}
It has been briefly stated, in the introductory section, that the standard Bohr Hamiltonian$^{21,22,24}$ which  is a $5-D$ space becomes the Davydov-Chaban Hamiltonian which is a $3-D$ space when  $\gamma$ is frozen. In the inverse square potential domain, the   Davydov-Chaban Hamiltonian  stated in Eq.(5) will be simply solved. The angular momentum in the intrinsic frame is written as$^{23}$ 

\begin{equation}
\hat{Q}^{2} = \dfrac{1}{\sin \theta} \dfrac{\partial}{\partial \theta} \sin \theta \dfrac{\partial}{\partial \theta} + \dfrac{1}{\sin^{2}\theta}\dfrac{\partial^{2}}{\partial \phi^{2}}.
\end{equation}
The eigenvalue equation as regards to the Hamiltonian is 
\begin{equation}
\hat{H}\Psi(\beta, \theta, \phi) =  E \Psi(\beta, \theta, \phi).
\end{equation}
By the usual method of separation of variable employed in some of our  quantum texts
\begin{equation}
 \Psi(\beta, \theta, \phi)= \chi(\beta) Y_{L,M}(\theta,\phi),
 \end{equation} 
where $Y_{L,M}(\theta,\phi)$ is the spherical harmonics and $\chi(\beta)$ is the radial part. The separated angular part obtained reads
\begin{equation}
-\left(\dfrac{1}{\sin \theta} \dfrac{\partial}{\partial \theta} \sin \theta \dfrac{\partial}{\partial \theta} + \dfrac{1}{\sin^{2}\theta}\dfrac{\partial^{2}}{\partial \phi^{2}} \right) Y_{L,M}(\theta,\phi)=L(L+1)Y_{L,M}(\theta, \phi),
\end{equation}
where $L$ is the angular momentum quantum number. After few steps, the simplified form of the radial part equation reads
\begin{equation}
\dfrac{d^{2}}{d\beta^{2}}\chi(\beta) + \dfrac{2}{\beta}\dfrac{d}{d\beta}\chi(\beta)-\dfrac{[L(L+1)+\beta_{0}]}{3\beta^{2}} \chi(\beta) +\epsilon\chi(\beta)=0,
\end{equation}
where $\dfrac{2B}{\hbar^{2}}\dfrac{\beta_{0}}{\beta^{2}}=v(\beta)$ has been considered as the reduced potential during the simplification and $\epsilon= \dfrac{2B}{\hbar^{2}}E$ is the reduced energy.

\subsection{Determination of the complete wave functions and the spectra ratios}
The methodology involves finding the complete wave function of the Eq.(10) and the eigenvalues equation is then obtained  from the wave function: these are quite easily done. In other to reduce the bulkiness and cumbersomeness that arise from the use of Nikiforov-Uvarov (NU) method$^{25}$ which has been employed in the literature and in ref$^{12}$ for solving similar differential equation,  Eq.(10) is easily solved using MAPLE software as in the refs$^{14,26,27}$ and the eigenfunctions obtained reads
 
\begin{equation}
\chi_{s,\nu,L,}(\beta) = \beta^{-1/2} \left[ C_{1,L}J_{\nu^{X(3)}}(\sqrt{\epsilon}\beta)+ C_{2,L}Y_{\nu^{X(3)}}(\sqrt{\epsilon}\beta)\right], 
\end{equation}
where $C_{1,L}$ and $C_{2,L}$ are the normalization constants associated with the Bessel functions of the first kind, $J_{\nu^{X(3)}}$, and second kind, $Y_{\nu^{X(3)}}$, respectively.  
In the domain of Eq.(4), the critical order associated with the $X(3)$ model in Eq.(10)  is
\begin{equation}
\nu^{X(3)}= \sqrt{\frac{L}{3}(L+1)+\beta_{0}+\dfrac{1}{4}}. 
\end{equation}

\noindent
If a boundary condition $\chi_{s,\nu^{X(3)},L}(\beta_{0})= 0$ is considered, then $C_{2,L, n}Y_{\nu^{X(3)}}(\sqrt{\epsilon}\beta)$ vanishes and the wave functions become 
\begin{equation}
\chi_{s,\nu^{X(3)}, L}(\beta) = \beta^{-1/2} \left[ C_{1,L}J_{\nu^{X(3)}}(\sqrt{\epsilon}\beta)\right].
\end{equation}   
By following the procedure for finding the eigenvalues written in the refs.$^{26,27}$, the  energy eigenvalues  in $\hbar=1$ unit reads:
\begin{equation}
\epsilon_{s, L, n_{\beta}} = 2n_{\beta}+1+\nu^{X(3)} = 2n_{\beta}+1+ \sqrt{\frac{L}{3}(L+1)+\beta_{0}+\dfrac{1}{4}}; \quad n_{\beta}=0,1,2,...
\end{equation}
From the $s-th$ zeros of the Bessel functions of order $\nu^{X(3)}$, the quantum number $s=n_{\beta}+1$. For the $X(3)$ model, the ground state energy levels are defined with $s=1$, the first $\beta$-excited levels are defined  with $s=2$ and  the second $\beta$-excited levels are defined  with $s=3$. There exist no $\gamma$-bands in the $X(3)$ model because $\gamma^{0}=0$. The $\epsilon_{s, L, n_{\beta}}$ can be reduced to $\epsilon_{s, L}$, consequently, the spectra ratios can be written as
\begin{equation}
R_{L_{s,n_{\beta}}}=\dfrac{\epsilon_{s,n_{\beta},L}-\epsilon_{1,0,0}}{\epsilon_{1,0,2}-\epsilon_{1,0,0}}.
\end{equation}
The normalization conditions for the wave function is the same condition as in refs$^{14,26,27}$, but it is worth noting that elemental volume is now proportional to the $\beta^{3}$ and not to the  $\beta^{4}$ in the standard Bohr-Mottelson space model.   The simplified normalization constants read
 \begin{equation}
       C_{1,L, n_{\beta}}= \left[ \sum_{n_{\beta}=0,1,2,3...} \dfrac{(\eta)_{n_{\beta}} \left(  \dfrac{\sqrt{\epsilon_{s, L, n_{\beta}}}}{2} \right)^{\xi-2} \beta_{0}^{(\xi)} }{n_{\beta}!\quad \xi \quad \left[ \Gamma \left( \dfrac{\xi}{2}\right) \right] ^{2} } \right] ^{-1/2},  
      \end{equation}     
where 
 \begin{equation}
 \xi=2\nu+2n_{\beta}+2,\quad \eta=2\nu^{X(3)}+n_{\beta}+1 \quad \mbox{and} \quad (\eta)_{n_{\beta}}=\eta(\eta+1)(\eta+2)...(\eta+n_{\beta}-1),\mbox{with} \quad (\eta)_{0}=1,   
  \end{equation} 
the complete wave functions for $X(3)$ with the one-parameter inverse square potential  model is
\begin{equation}
\chi_{s,\nu^{X(3)},L,n_{\beta}}(\beta)=  \left[ \sum_{n_{\beta}=0,1,2,3...} \dfrac{(\eta)_{n_{\beta}} \left(  \dfrac{\sqrt{\epsilon_{s, L, n_{\beta}}}}{2} \right)^{\xi-2} \beta_{0}^{(\xi)} }{n_{\beta}!\quad \xi \quad \left[ \Gamma \left( \dfrac{\xi}{2}\right) \right] ^{2} } \right] ^{-1/2}    \beta^{-1/2} J_{\nu^{X(3)}}(\sqrt{\epsilon_{s, L, n_{\beta}}}\beta).
\end{equation} 

\section{$B(E2)$ transition rates}
The general electric quadrupole operator is written as$^{28}$  
\begin{equation}
T^{E2}_{\mu}=t\beta\left[ D_{\mu,0}^{(2)}(\theta_{i}) \cos \gamma  + \dfrac{1}{\sqrt{2}} \left(D_{\mu,2}^{(2)}(\theta_{i}) + D_{\mu,-2}^{(2)}(\theta_{i})\right) \sin  \gamma \right], 
\end{equation}
where $D(\theta_{i})$ are the Wigner functions of the Euler angle and $t$ is known as a scale factor. 
For $\gamma^{0}=0$,  
\begin{equation}
T^{E2}_{\mu}= t\beta\sqrt{\dfrac{4\pi}{5}} Y_{2\mu}(\theta,\phi). 
\end{equation}
The $B(E2)$ is written as   
\begin{equation}
B(E2;sL\longrightarrow s^{\prime} L^{\prime}) =\dfrac{1}{2sL+1}|\left\langle s^{\prime}L^{\prime}||T^{E2}|| sL\right\rangle |^{2} =t^{2} \left(C_{L0,20}^{L^{\prime}0} \right)^{2} I^{2}_{sL;s^{\prime}L^{\prime}},
\end{equation}

where the coefficients, $C_{L0,20}^{L^{\prime}0}$ are the Clebsch-Gordan coefficients$^{29}$, and 

\begin{equation}
I_{sL;s^{\prime}L^{\prime}}= \int_{0}^{\beta_{0}} \beta \chi_{s,\nu,L,n_{\beta}}(\beta) \chi_{s^{\prime},\nu^{\prime},L^{\prime},n^{\prime}_{\beta}}(\beta)\beta^{2}d\beta , 
\end{equation}
are the integrals over $\beta$.

\section{Discussion of the analytical results, the  numerical results, and the experimental realization of the model}
\subsection{The theoretical assessment of the present model and the $X(5)$ model }
There is need to discuss and  examine if the present model presents a dynamical symmetry: that is if, there is a link between the present model and the  nuclei at the corners of the Casten triangle$^{49}$ via the variatinal procedure. There is also a need to examine the roles of the nature of the potential, Eq.(4), in the critical point symmetries (CPS) scheme. There is a need to clarify if the potential provides an improved solutions compared to the other solutions found in the literature. Most importantly, the similarities between the present model and the $X(5)$ model should be analytically discussed.  These assessments  are done via the presentation, comparison and the discussion of the numerical solutions for some quadrupole collective signatures, such as the spectra ratios for some quantum levels and the $B(E2)$ transition probabilities for the available states. The $\gamma$ staggering effect which is one of the quadrupole collective signatures is not considered, since it does not exist in $X(3)$ model because $\gamma$ is frozen to zero. 

\noindent
Some important primary solutions for the  collective model of Eq.(5)  and such Hamiltonian from the various potentials for $X(5)$ models$^{2,27,32}$ and others, are the normalized wave functions from which the $B(E2)$ transition probabilities are determined and the energies of the quantum levels. Albeit, the complete wave functions  and the energies of the $X(5)$ model contain the $\gamma$-part which is essential in the computation of the $\gamma$-excited states, the $\beta$-part wave functions of  the $X(5)$ and $X(3)$ are in the form of Bessel functions having a critical order, $\nu$, with a preferred condition $-(2n_{\beta}+1)<\nu<-2n_{\beta}$ leaving the  Bessel functions with $4n_{\beta}$ complex roots of which none is imaginary$^{26}$: this condition is used to determine the continuous spectra equation for the two models. For instance, Eq.(14) is  similar to the spectra equation obtained  in the $\beta$-part of  $X(5)$  model in refs$^{2,27,32}$, the difference being the critical order. In ref.$^{27}$ where the same potential is employed,  the critical order is  

\begin{equation}
\nu^{X(5)}= \sqrt{\frac{L}{3}(L+1)+\beta_{0}+\dfrac{9}{4}}.
\end{equation}
\noindent
Now that it has been established that both the $X(3)$ model and the $X(5$) model have their  critical orders, $\nu(L,\beta_{0})$ which dominates the description of the energy spectra in the ground states and the $\beta$-excited states, from their  Bessel functions, there is a need to compare the two critical orders.
\noindent
Firstly, in the comparison of the Eq.(12) and Eq.(23) and from their numerical values computed in Table 1. for $L=0$ to $L=10$ at different values of $\beta_{0}$,  it can be deduced  that 
\begin{equation}
 \nu^{X(3)}(\beta_{0}=c+2)=\nu^{X(5)}(\beta_{0}=c); \quad c=0,1,2,...  
 \end{equation}  
In both cases, critical orders increase with increase in the angular momentum, $L$, and with increase in the variation parameter, $\beta_{0}$. These effects of  $L$ and $\beta_{0}$ in $\nu$ are also seen in the numerical values of  the energies in all the levels. 

\noindent 
Secondly, the exact relationship between the $\nu^{X(3)}$ and the $\nu^{X(5)}$ stated in Eq.(24) does not reflect in the exact comparison of their total energies of the levels,  that is   
\begin{equation}
\epsilon^{X(3)}(\beta_{0}=c+2)\neq\epsilon^{X(5)}(\beta_{0}=c),
\end{equation}
because the total energy of the $X(5)$ contains the $\gamma$-part solutions. It is worth noting that in the ground states and the $\beta$-excited states, the  relation
\begin{equation}
\epsilon_{gs,L}=2+\epsilon_{\beta_{1},L}=4+ \epsilon_{\beta_{2},L},
 \end{equation}
holds in all the levels for both $X(3)$ and  $X(5)$. This third remark is shown from the numerical values in the Table 2. The visual  lines of the energies in the three states: the ground state and the $\beta$-excited states at $\beta_{0}=2$ are plotted and shown  in Figure 1(a). In all the three states, for the same quantum levels, energy values for the $X(3)$ are higher than the energies  of  the $X(5)$.

\noindent
Another significant remark deduced from the numerical values of $\nu$, tabulated in Table 1., is that  $\nu^{X(5)}$  at $L=0$, correspond to $\nu^{X(3)}$ at $L=2$: the visual lines are seen in Figure 1(b) where the brown dotted line of $\nu^{X(3)}$ for different values of  $\beta_{0}$ lies on the green line of $\nu^{X(5)}$  at $L=0$ for different  values of  $\beta_{0}$. The plot shows that no other lines of $\nu^{X(3)}$ coincides with the $\nu^{X(5)}$ again, they only increase with the increase in the angular momentum, $L$. Analytically, it can be stated that
\begin{equation}
\nu^{X(5)}(L=0)= \nu^{X(3)}(L=2) \quad \forall \quad \beta_{0}.
\end{equation}
\noindent
At, this juncture, it is wise to test for the stationary properties of the potential parameters as to know which of the parameters of the potential plays  a major role in the dynamical symmetry of the model. It is expected that the n-derivatives of $\nu$ with respect to such parameter  should be continuous in all the quantum levels. Having known the values of such parameter for  certain isotopes, the numerical values for the spectra ratios, the $B(E2)$  and other nuclei collective signatures can be computed. Luckily enough in the present work, the potential employed  is a one parameter dependent potential, it depends solely on the $\beta_{0}$. Therefore, the derivatives of $\nu$ with respect to the $\beta_{0}$ are taken and the continuous phenomenon and the lines of the $\frac{\delta^{2} \nu}{\delta \beta_{0}^{2}}$ and the $\frac{\delta \nu}{\delta \beta_{0}}$ at $L=2,4$ and at $L=6$ are respectively shown in  Figures 1(c) and 1(d). Since the derivatives of $\nu$ with respect to $\beta_{0}$ is continuous, $\beta_{0}$ will yield to the variation procedure. Its forward or backward variation will provide some physical significance in the interpretation of the solution within the context of  phase shape transition. \\

\noindent
The spectra ratios of all the bands are normalized to the lowest excited energy level, $2_{1,0}$. The numerical values are computed and tabulated in Table 3. The ground state spectra ratios  at  different values of $\beta_{0}=0,1,2,3,4,5, 15$ and $\beta_{0}=\infty$,  are compared with  the $X(5)$ in ref.$^{27}$ where the same potential has been used. The ground state spectra ratios and the $\beta$-excited spectra ratios at  different values of $\beta_{0}=0,1,2,3,4,5, 15$ and $\beta_{0}=\infty$,  are compared with the $U(5)$ in ref.$^{29}$, the $X(5)$ model in ref.$^{2}$ and the $SU(3)$ reported in ref.$^{30}$. The  spectra ratios at $\beta_{0}=0$ tend to the $U(5)$ vibrational limit while the  spectra ratios at $\beta_{0}=\infty$ tend to the $SU(3)$ rotational limit.  It can be seen, in the ground state where $SU(3)$ data are available, that  $X(3)(\beta_{0}= \infty)$ and $X(5)(\beta_{0}= \infty)$, both with $R_{4/2}=3.296$ and  marked with a  `$\dagger$' sign, approach $SU(3)$ whose `rotational' excitation signature$^{1}$, $R_{4/2}=3.333$. The solutions from $\beta_{0}=0$ to $\beta_{0}=\infty$ spread over the $X(5)$. That is, the solutions move to the $X(5)$ as $\beta_{0}$ leaves zero and leave $X(5)$ going close to $SU(3)$ as $\beta_{0}$ goes to $\infty$, in a forward manner. The lines plotted in  Figures 2(a), 2(b) and 2(c) visually show these comparisons in the ground state and the $\beta$-excited states respectively. In  Figure 2(a), the width of the spread of the solutions over the $X(5)$ line is indicated by the arrow range line between $\beta_{0}=0$ and $\beta_{0}=\infty$. The `nature' of critical point symmetry transitions for different isotopes, constrained to one-parameter potentials, can be investigated using a variational technique. This technique was introduced in the ref.$^{32}$, used in ref.$^{12}$  to retrieve  $U(5)$ and $O(6)$ in the ground state  from the $E(5)$ within the domain of the one-parameter inverse square potential, it is also used in refs$^{12,14,26,27}$  and others. The forward variation of the `control parameter', $\beta_{0}$,  causes the nuclei transition from $X(5)$ to $SU(3)$.  The  nuclear shape phase region under investigation can predict the directions of the variation: whether  forward variation or backward variation and  also depends on the potential's boundary conditions. The rate of change of the spectra ratios is maximized for each $L$ by using this technique. Each angular momentum is considered and treated separately in terms of the variation parameter, $\beta_{0}$, as the critical values of the spectra ratios are distinct. As shown in Table 4., the distinct value of  $\beta_{0}$ that corresponds to each angular momentum obtained via $\beta_{0}$-optimization scheme$^{12,14,26,27}$ are labeled $\beta_{0,max}$. The calculated values of the $X(3)$ as a function of $\beta_{0,max}$,  labeled $X(3)$-var are compared with $X(3)$-IW models in ref.$^{8}$, $X(3)-\beta^{6}$ model found in ref.$^{10}$ and $X(3)$-D model in ref.$^{12}$. There is a strong agreement between the present $X(3)$-var and $X(3)$-IW, the agreement is moderate between $X(3)-\beta^{6}$ model, $X(3)$-D model and the preset $X(3)$ model. In levels $0_{1,0}$ and $2_{1,0}$, $\beta_{0,max}=\beta_{0}$ because, any value of  $\beta_{0}$ chosen will yield $0.000$ and $1.000$ for the calculations of $0_{1,0}$ and $2_{1,0}$ respectively. $\beta_{0,max}$ increases with increase in the angular momentum and its values  are obtained at the points where the increase in $\beta_{0}$ becomes steep from the procedure $\dfrac{d}{d\beta_{0}}R_{L/2}\vert_{\beta_{0}=max}$, so that the rate of change of the spectra ratio is maximized. This procedure is carried out because, the shape formation of the critical points changes swiftly: such rapid change is seen just in the solutions tabulated in the third section of the Table 3., where the $U(5)$ vibrational limit properties of the solutions at $\beta_{0}=0$  rapidly change as $\beta_{0}>0$. From the numerical values of the spectra ratios computed at   $\beta_{0}=0$, in Table 3., the relation   
\begin{equation}
R_{L/2}(gsb)=2+R_{L/2}(\beta_{1})=4+ R_{L/2}(\beta_{2}),
 \end{equation}
can be deduced. Regardless of the fractional part, this relation agrees with the $U(5)$ vibrational limit$^{29}$.  This also suggests that the solutions at $\beta_{0}=0$ approach the $U(5)$ vibrational limit since the relation is not observed for spectra ratios  at other values of $\beta_{0}>0$ .  This is an observable effect or a signature from Eq.(26) while  the effect of Eq.(24) is  observed in the comparison of  spectral ratios of $X(3)$ and  $X(5)$ such that
 \begin{equation}
 R_{L/2}^{X(3)}(\beta_{0}=c+2)= R_{L/2}^{X(5)}(\beta_{0}=c); \quad c=0,1,2,... \quad .
 \end{equation} 
 
\noindent
The $B(E2)$ ratio is an additional structural signature for the quadrupole collective states that has been used in the searching for the shape phase change and for the critical point between two fixed symmetries. It is determined primarily from the normalized wave functions of Eq.(18) in Eq.(21). The numerical values of the $B(E2)$ transition rates of the present $X(3)$  model at $\beta_{0} = 0,1, 2$ and  $\beta_{0}=\infty$, normalized to the  $B(E2;2_{1,0}\rightarrow 0_{1,0})=100$ units are tabulated in the first section of Table 5. and are compared with the $U(5)$ in ref.$^{29}$, $X(5)$ in ref.$^{2}$ and $SU(3)$ reported in ref.$^{30}$. The solutions at $\beta_{0}=0$ lie close the  $U(5)$ and go close to the axially symmetric prolate rotor, $SU(3)$, at $\beta_{0}=\infty$. As expected, the $B(E2)$ values decrease with the increase in the $\beta_{0}$ and the solutions spread over  $X(5)$. The visual comparisons in the ground state and the $\beta$-excited states are respectively shown in Figures 2(d), 2(e) and 2(f). In the  second section of  Table 5., the $B(E2)$-var calculated in the present model, are compared with the  $B(E2)$ solutions obtained from the infinite square well potential, $X(3)$-IW model, in ref$^{8}$: the agreements between the two models in all the states are strong. This is because, the inverse square potential which can also be referred to as inverse square well has most properties of the infinite square well. Since each value of $\beta_{0,max}$ is distinct for each angular momentum, the values of  $\beta_{0,max}^{(i)}$ and $\beta_{0,max}^{(f)}$ are the  $\beta_{0,max}$ which correspond to the initial and final angular momenta considered in the transitions between the levels. They are obtained from the optimization procedure done for the spectra ratios recorded in the Table 4.

\noindent
Before we discuss the experimental realization of this model, there is a need to show if the solutions presented and discussed before now, are dynamical symmetry. Now considering the interpretation of the Casten triangle$^{49}$ where the rotational excitation signature, $4_{1,0}$, of the isotopes mapped along $U(5)\rightarrow SU(3)$ region increase from $2.000$ to $3.333$, if either the model condition

\begin{equation}
U(5)\leq X(3) < SU(3),  \quad \forall \quad \beta_{0}, 
\end{equation}
is satisfied for both the spectra ratios and the $B(E2)$ or the numerical condition for the spectra ratios
\begin{equation}
\dfrac{L+2}{2} \leq R_{L+2/2_{1,0}}(X(3))< \dfrac{(L+2)(L+3)}{6}, 
\end{equation}
is satisfied, then the solutions are said to be dynamical symmetry since the spectra ratios of the  $U(5)$ is the same as the left hand side algorithm of Eq.(31) while  $SU(3)$ obeys  the right hand side algorithm of the same Eq.(31). The entire solutions from $\beta_{0}=0$ to $\beta_{0}=\infty$, in the Table 3. and Table 5., satisfy the conditions. $X(5)$ models also satisfy these conditions.

\subsection{The experimental realization of the model } 
In the prediction of spectra ratios of the present $X(3)$ model, twelve best critical point  isotopes, $^{102}$Mo ref.$^{33}$, $^{104-108}$Ru  chain in refs$^{34-36}$,  $^{120-126}$Xe chain in refs$^{37-40}$,  $^{148}$Nd in ref.$^{41}$, $^{184-188}$Pt chain in refs$^{42-44}$ are considered for the experimental realization and for the numerical application. Since the model is meant to be compared with the $X(5)$ critical symmetry,  $N=90$ isotones: $^{150}$Nd in ref.$^{45}$, $^{154}$Gd in ref.$^{46}$ and $^{156}$Dy in ref.$^{47}$ known to be  $X(5)$ candidates, are compared with the $X(3)$ theoretical model. In order to obtain the quality factor of the agreement, Eq.(14) is fitted with the experimental spectra and the free parameter of the potential, $\beta_{0}$, is determined for each isotope being considered. Having known the value of $\beta_{0}$ for each isotope, the quality factor which is the quantity that measures the agreement is  

\begin{equation}
\sigma=\sqrt{\dfrac{\sum_{i}^{m}[(R_{s,L})^{Exp}_{i}- (R_{s,L})^{Theor}_{i}]^{2}}{m-1}},
\end{equation}

\noindent
where $m$ is the number of available experimental states, $(R_{s,L})^{Exp}_{i}$ and $(R_{s,L})^{Theor}_{i}$ represent the experimental and the theoretical spectral ratios of the $i^{th}$ levels normalized to the ground state. These comparisons between the experimental spectra ratios  and the theories are shown in the Table 6. While using the isotopes with the smallest quality factor such as $^{108}$Ru,  $^{108}$Ru, $^{122}$Xe, $^{124}$Xe, $^{126}$Xe to judge the agreement, it must be noted that isotopes with highest experimental states will yield smallest root mean square values. Regardless, except for $^{188}$Pt with the highest quality factor, the quality factors for all the isotopes considered are in moderate agreement with the experimental data. This shows that the $X(3)$ inverse square model is a critical point symmetry model which presents  the transition `around' or near the vibrational $U(5)$ to prolate axially deformed shapes, $SU(3)$. In the same manner, $^{150}$Nd, $^{154}$Gd  and $^{156}$Dy are compared with the present $X(3)$ model in third section of Table 6. and the visual line in the ground state and the $\beta$-excited states are shown in Figure 3. There is no rational agreement  in the ground states for the three isotopes,  but an excellent prediction appears in the first $\beta$-excited state. This suggests that this present model can compensate the $X(5)$ models whose  predictions are excellent in the ground states but moderately bad in the  $\beta$-excited states. 
The  ground states and the first $\beta$-excited states of the $B(E2)$  data for  $^{102}$Mo, $^{104}$Ru, $^{108}$Ru, $^{120}$Xe, $^{122}$Xe, $^{124}$Xe,  $^{148}$Nd and one $X(5)$ candidate: $^{152}$Sm isotope in ref.$^{48}$ are placed for comparison with the present $X(3)$ model. The $B(E2)$ values in Table 7. are not fitted, they are calculated with the values of $\beta_{0}$ recorded for the same isotopes from their fits on spectra. For $^{152}$Sm, the $\beta_{0}=1.052$ and the quality the factor $\sigma=0.357$.

\section{Conclusion}
A new critical point symmetry model (CPSM) which is  the $X(3)$ with the inverse square potential which presents  the change from the region `around' the $U(5)$ vibrational limit  to prolate axially deformed shapes, $SU(3)$ limit in the frame work Bohr-Mottelson Hamiltonian is  presented. The analytical form of the wave function and the one parameter-dependent energy spectra are obtained. The model is compared with the $X(5)$ model analytically to show the depth in the similarities and discrepancies in the context of the comparison, some new set of useful equations Eq.(24)- Eq.(31) are deduced. The model has proven sufficiently helpful in the description of $U(5)\rightarrow SU(3)$ region of the Casten triangle$^{49}$, since the solutions satisfy Eq.(30) and Eq.(31). The comparison of the present model with those found in the literature such as $X(3)$-IW model$^{8}$, $X(3)-\beta^{6}$ model$^{10}$ and $X(3)$-D model$^{12}$ is good. There is a strong agreement between the present $X(3)$-var and $X(3)$-IW model$^{8}$ because the inverse square potential, its properties, are similar to those of infinite square well potential. The comparisons of the model with twelve best critical point  isotopes, $^{102}$Mo ref.$^{33}$, $^{104-108}$Ru  chain in refs$^{34-36}$,  $^{120-126}$Xe chain in refs$^{37-40}$,  $^{148}$Nd in ref.$^{41}$, $^{184-188}$Pt chain in refs$^{42-44}$ are moderate. An excellent agreement which appears in the first $\beta$-excited state in the comparison of the $X(3)$ via the inverse square potential with three  $N=90$ isotones: $^{150}$Nd in ref.$^{45}$, $^{154}$Gd in ref.$^{46}$ and $^{156}$Dy in ref.$^{47}$ known to be  $X(5)$ candidates,  suggests that this present model can compensate the $X(5)$ models whose predictions are very good in the ground states but performed poorly in the  $\beta$-excited states.

\section*{Author Contribution:}\begin{small}
 KRA and KJO conceived, designed and drafted the work. KRA solved the differential equations, determined the wave functions, eigenvalues, spectra ratios, carried out variational technique to maximize the spectra ratios,  and obtained the expression for the $B(E2)$ transition probabilities while KJO confirmed the correctness of the solutions and provided their numerical values.  KRA plotted the graphs and LaTeX the work while KJO proof-read and approved the submission.

\end{small}
\section*{Data availability statement}
\begin{small}
The experimental data used for the experimental realization of our model are the published data: they are being cited and referenced accordingly. They are also available in the nuclear data sheet  repository web link as of 2022:  https://www.nndc.
bnl.gov/nudat3/chartNuc.jsp.
\end{small}
\section*{Funding Information}
No funding of any form is received for the course of this work.

 \subsection*{{\Large References}} 
     
 \begin{description}

\begin{footnotesize}
\item[][1]  Iachello, F. \textit{Analytic Description of Critical Point Nuclei in a Spherical-Axially Deformed Shape Phase Transition.} Phys. Rev. Lett. 87, 052502. doi: https://doi.org/10.1103/PhysRevLett.87.052502 (2001).
\item[][2] Iachello, F. \textit{Dynamic Symmetries at the Critical Point.} Phys. Rev. Lett. 85, 3580.\\ doi: https://doi.org/10.1103/PhysRevLett.85.3580 (2000).
\item[][3] Iachello, F. and   Arima, A. \textit{The Interacting Boson Approximation Model.} Cambridge University Press, Cambridge.\\ doi: https://doi.org/10.1017/CBO9780511895517 (1987).

\item[][4]  Bohr, A. and  Mottelson, B. \textit{The structure of Angular Momentum in Rapidly Rotating Nuclei.} Nuclear Physics A., \textbf{354},  (1-2),  303-31. doi: https://doi.org/10.1016/0375-9474(81)90604-7 (1981).
\item[][5] Bohr, A. and  Mottelson, B. \textit{ Nuclear Structure Vol. I: Single-Particle Motion} World Scientific Publishing Co. Pte. Ltd.  Singapore. url: http://nuclphys.sinp.msu.ru/books/b/Bohr-Mottelson-I.pdf (1975).
\item[][6] Bohr, A. \textit{The Coupling of Nuclear Surface Oscillations to the Motion of Individual Nucleons.}
 Mat. Fys. Medd. Dan. Vid. Selsk. 26, 14. url: http://www.xuantianlinyu.com.cn/Jabref/RefPdf/Bohr1952pp.pdf(1952).
\item[][7]  Wilets, L. and   Jean, M. \textit{Surface Oscillations in Even-Even Nuclei.} Phys. Rev. C 102, 788.\\ doi: https://doi.org/10.1103/PhysRev.102.788 (1956).
\item[][8] Bonatsos,  D.,  Lenis, D.,  Petrellis, D.,  Terziev, P.A. and  Yigitoglu, I.  \textit{$X(3)$: an exactly separable $\gamma$-rigid version of the $X(5)$ critical point symmetry.}  Physics Letters B, \textbf{632}, 238-242. doi: http://dx.doi.org/10.1016/j.physletb.2005.10.060 (2006).
\item[][9] Budaca, R. Quartic oscillator potential in the $\gamma$-rigid regime of the collective geometrical model, Eur. Phys. J. A., \textbf{50}, 87. doi: https://doi.org/10.1140/epja/i2014-14087-8 (2014).
\item[][10] Budaca, R. \textit{Harmonic Oscillator Potential with a Sextic Anharmonicity
in the Prolate $\gamma$-rigid Collective Geometrical Model.} Physics Letters B, \textbf{739}, 56-61. doi: http://dx.doi.org/10.1016/j.physletb.2014.10.031 (2014).
\item[][11]  Alimohammadi, M. and Hassanabadi, H. \textit{The $X(3)$ Model for the Modified Davidson
Potential in a Variational Approach.}  International Journal of Modern Physics E. \textbf{26}(9), 1750054. doi: http://dx.doi.org/10.1142/S0218301317500549 (2017).
\item[][12] Yigitoglu, I. and  Gokbulut, M. \textit{Bohr Hamiltonian for $\gamma=0^{0}$ with Davidson Potential.}  Eur. Phys. J. Plus, \textbf{132}, 345. doi: http://dx.doi.org/10.1140/epjp/i2017-11609-3 (2017).
\item[][13] McCutchan, E. A., Bonatsos, D., and Casten, R. F.  \textit{Connecting the $X(5)-\beta^{2}$, $X(5)-\beta^{4}$, and $X(3)$ Models to the Shape-Phase Transition Region of the Interacting Boson Model}. HNPS Advances in Nuclear Physics, 15, 118-127. doi: https://doi.org/10.12681/hnps.2628 (2020).
\item[][14] Ajulo, K.R. and  Oyewumi, K.J. \textit{Symmetry Solutions at $\gamma^{0}=\pi/6$ for Nuclei Transition Between $\gamma^{0}=0$ and $\gamma^{0}=\pi/3$ Via a Variational Procedure.}  Physica Scripta, \textbf{137}(90). doi: https://doi.org/10.1088/1402-4896/ac76ed (2022).
\item[][15]   Davidson, P.M. \textit{Eigenfunctions for Calculating Electronic Vibrational Intensities.}  Proc. R. Soc. London, Ser. A., \textbf{135}, 459. doi; https://doi.org/10.1098/rspa.1932.0045 (1932).
\item[][16]   Bonatsos \textit{et al}., \textit{Exactly Separable Version of the Bohr Hamiltonian with the Davidson Potential.} Phys. Rev. C 76, 064312. doi: https://doi.org/10.1103/PhysRevC.76.064312 (2007).
\item[][17] Kratzer, A.  Die ultraroten Rotationsspektren der Halogenwasserstoffe. Z. Physik \textbf{3}, 307. \\doi: https://doi.org/10.1007/BF01327754 (1920).
\item[][18]  Fortunato, L. \textit{Solutions of the Bohr Hamiltonian, a compendium.} Eur. Phys. J. A.,  \textbf{26} (1), 1-30.\\ doi: https://doi.org/10.1140/epjad/i2005-07-115-8 (2005).
\item[][19]   Boztosun, I.,  Bonatsos, D. and  Inci, I.  \textit{Analytical Solutions of the Bohr Hamiltonian with the Morse Potential.} Phys. Rev. C 77, 044302. doi: https://doi.org/10.1103/PhysRevC.77.044302 (2008).
\item[][20]  Inci, I.,   Bonatsos, D. and  Boztosun, I.  \textit{Electric Quadrupole Transitions of the Bohr Hamiltonian with the Morse Potential.} Phys. Rev. C., \textbf{84}, 024309. doi: https://doi.org/10.1103/PhysRevC.84.024309 (2011).
\item[][21] Bohr,  A. and  Mottelson, B.  \textit{Collective and Individual-Particle Aspects of Nuclear Structure.}  Mat-Fys. Medd. \textbf{27}(16). 1-174. url: https://cds.cern.ch/record/213298/files/p1.pdf (1953).
\item[][22] Bohr, A. and  Mottelson, B. \textit{Nuclear Structure, Vol. II: Nuclear Deformations}.  W. A. Benjamin, Inc., Reading, Massachusetts,  \textbf{748},  37-50. url: http://nuclphys.sinp.msu.ru/books/b/Bohr-Mottelson-I.pdf (1975).
\item[][23]  Davydov, A. S. and  Chaban, A. A.  \textit{Rotation-Vibration Interaction in Non-Axial Even Nuclei.}  Nucl. Phys. \textbf{20}, 499-508. doi; https://doi.org/10.1016/0029-5582(60)90191-7 (1960).
\item[][24] Bohr, A.  \textit{The Coupling of Nuclear Surface Oscillations to the Motion of Individual Nucleons.}  Dan. Mat. Fys. Medd. \textbf{26}(14). http://www.xuantianlinyu.com.cn/Jabref/RefPdf/Bohr1952pp.pdf (1952).
\item[][25] Nikiforov, A.V. and Uvarov, V.B.  \textit{Special Functions of Mathematical Physics.}  Birkhauser, Bassel.\\ doi: http://dx.doi.org/10.1007/978-1-4757-1595-8 (1988).
\item[][26]  Ajulo, K.R., Oyewumi, K.J., Oyun, O.S. and Ajibade, S.O. \textit{$U(5)$ and $O(6)$ Shape Phase Transitions Via $E(5)$ Inverse Square Potential Solutions.}   Eur. Phys. J. Plus, 136(500). doi: https://doi.org/10.1140/epjp/s13360-021-01451-7 (2021).
\item[][27] Ajulo, K.R., Oyewumi, K.J., Oyun, O.S. and Ajibade, S.O. \textit{$X(5)$ Critical Symmetry with Inverse Square Potential Via a Variational Procedure.}  Eur. Phys. J. Plus \textbf{137}(90). doi: https://doi.org/10.1140/epjp/s13360-021-02276-0 (2022).
\item[][28]  Casten, R.F. and   Zamfir, N.V. \textit{Empirical Realization of a Critical Point Description in Atomic Nuclei.} Phys. Rev. Lett. 87, 052503. doi: https://doi.org/10.1103/PhysRevLett.87.052503 (2001).
\item[][29]  Rowe, D.J.,  Turner, P.S. and   Repka, J. \textit{Spherical Harmonics and Basic Coupling Coefficients for the Group $SO(5)$ in an $SO(3)$ Basis.}  J. Math. Phys. 45, 2761. doi: https://doi.org/10.1063/1.1763004 (2004).
\item[][30]  Bonatsos, D.,   Lenis, D.,   Minkov, N.,  Raychev, P.P. and  Terziev. P. A. \textit{Extended $E(5)$ and $X(5)$ Symmetries: Series of Models Providing Parameter-Independent Predictions}. Physics of Atomic Nuclei, \textbf{67}(10), 1767-1775.\\ doi: https://doi.org/10.1134/1.1811176 (2004).
\item[][31]  Kotb. M.  \textit{$U(5)-SU(3)$ Nuclear Shape Transition Within the Interacting Boson Model Applied to Dysprosium Isotopes.} Physics of Particles and Nuclei Letters, \textbf{13}(4), 451-459. doi: 10.1134/S1547477116040075 (2016).
\item[][32]  Bonatsos, D.,  Lenis, D.,  Minkov, N.,  Raychev, P.P. and   Terziev. P.A.  \textit{Ground State Bands of the $E(5)$ and $X(5)$ Critical Symmetries Obtained from Davidson Potentials Through a Variational Procedure.} Physics Letters B, 584, 40-47.\\ doi:10.1016/j.physletb.2004.01.018 (2004).
\item[][33] DE Frenne, D.   Nucl. Data Sheets, \textbf{110}(8),  1745-1915. doi: https://doi.org/10.1016/j.nds.2009.06.002 (2009).
\item[][34] Blachot, J.    Nucl. Data Sheets, \textbf{108}(10),  2035-2172.
 doi: https://doi.org/10.1016/j.nds.2007.09.001 (2007).
\item[][35] DE Frenne, D.  and  Negret, A.   Nucl. Data Sheets, \textbf{109}(4), 943-1102.
  doi: https://doi.org/10.1016/j.nds.2008.03.002 (2008).
\item[][36]  Blachot, J.    Nucl. Data Sheets, \textbf{91}(2), 135-296.  doi: https://doi.org/10.1006/ndsh.2000.0017(2000).
\item[][37] Kitao, K.,  Tendow, Y. and  Hashizume, A.   Nucl. Data Sheets, \textbf{96}(2), 241-390. doi: https://doi.org/10.1006/ndsh.2002.0012 (2002).
\item[][38] Tamura, T.   Nucl. Data Sheets, \textbf{108}(3),  455-632. doi: https://doi.org/10.1016/j.nds.2007.02.001(2007).
\item[][39]  Katakura, J.  and  Wu, Z.D.   Nucl. Data Sheets, \textbf{109}(7),  1655-1877. 
 doi: https://doi.org/10.1016/j.nds.2008.06.001 (2008).
\item[][40] Limura, H., Katakura, J. and Ohya, S.  Nucl. Data Sheets, 180, 1-413. doi: https://doi.org/10.1016/j.nds.2022.02.001 (2022).
\item[][41] Nica, N.    Nucl. Data Sheets 117, 1-229. doi: https://doi.org/10.1016/j.nds.2014.02.001 (2014).
\item[][42] Baglin, C.M.    Nucl. Data Sheets,  \textbf{111}(2), 275-523. doi: https://doi.org/10.1016/j.nds.2010.01.001 (2010).
\item[][43]  Baglin, C.M.    Nucl. Data Sheets,  \textbf{99}(9), 1-196. doi: https://doi.org/10.1006/ndsh.2003.0007 (2003).
\item[][44] Singh, B.  Nucl. Data Sheets, \textbf{95}(2), 387-541. doi: https://doi.org/10.1006/ndsh.2002.0005 (2002).
\item[][45]  Basu, S.K. and Sonzogni, A.A.  Nucl. Data Sheets, \textbf{114}(4-5), 435-660. https://doi.org/10.1016/j.nds.2013.04.001 (2013).
\item[][46] Reich, C.W.  \textit{Nuclear data sheets for A = 154}. Nucl. Data Sheets, \textbf{110}, 2257. https://doi.org/10.1016/j.nds.2009.09.001 (2009).
\item[][47] Reich, C.W.  Nuclear Data Sheets for $A = 156$ Nucl. Data Sheets, \textbf{113}(11),  2537-2840.\\ https://doi.org/10.1016/j.nds.2012.10.003 (2012).
\item[][48] Martin, M.J.  Nucl. Data Sheets, \textbf{114}(11),  1497-1847. https://doi.org/10.1016/j.nds.2013.11.001 (2013).
\item[][49] R.F. Casten, \textit{Nuclear Structure from a Simple Perspective.} Oxford University Press, Oxford.\\ url: https://radium.phys.uoa.gr/ebk/Casten-NuclearStructureFromASimplePerspective.pdf (1990).

\end{footnotesize}
\end{description}

\begin{table}[hbtp]
\caption{{\small The comparison between the  calculated  values  of  the critical orders, $\nu$, for the $X(5)$ in Eq.(23) and $X(3)$ in Eq.(12) for both the even and the odd values of $\beta_{0}$.}}
\begin{small}
\begin{center}
\begin{tabular}{ccccccccccccc}
\hline 
$L$ &$\beta_{0}=0$&&$\beta_{0}=2$&&$\beta_{0}=4$&&$\beta_{0}=6$&&$\beta_{0}=102$&$\beta_{0}=100$\\
 
\hline
&$X(3)$&$X(5)$&$X(3)$&$X(5)$&$X(3)$&$X(5)$& $X(3)$ &$X(5)$&$X(3)$&$X(5)$\\
\hline
0&0.500&1.500&1.500 &2.062&2.062&2.500&2.500&2.062	&10.112&10.112 \\
2&1.500&2.062&2.062 &2.500&2.500&2.872&2.872&2.500	&10.210&10.210 \\
4&2.630&2.986&2.986 &3.304&3.304&3.594&3.594&3.304	&10.436&10.436 \\
6&3.775&4.031&4.031 &4.272&4.272&4.500&4.500&4.272	&10.782&10.782 \\
8&4.924&5.123&5.123 &5.315&5.315&5.500&5.500&5.315	&11.236&11.236 \\
10&6.076&6.238&6.238&6.397&6.397&6.551 &6.551&6.397	&11.786& 11.786\\
\hline	
 &$\beta_{0}=1$&&$\beta_{0}=3$&&$\beta_{0}=5$&&$\beta_{0}=7$&&$\beta_{0}=101$&$\beta_{0}=103$\\
\hline
0&1.118&1.803&1.803 &2.291&2.291&2.693&2.693&3.041	&10.062&10.259 \\
2&1.803&2.291&2.291 &2.693&2.693&3.041&3.041&3.354	&10.161&10.356 \\
4&2.814&3.149&3.149 &3.452&3.452&3.731&3.731&3.990	&10.388&10.579 \\
6&3.905&4.153&4.153 &4.387&4.387&4.610&4.610&4.823	&10.735&10.920 \\
8&5.025&5.220&5.220 &5.408&5.408&5.590&5.590&5.766	&11.191&11.369 \\
10&6.158&6.318&6.318 &6.474&6.474&6.627&6.627&6.776	&11.747&11.913\\
\hline	

\end{tabular}
\end{center}
\end{small}
\end{table}

\begin{table}[hbtp]
\caption{{\small The ground state  and the $\beta$ exited states energies in $\hbar =1$ unit,  denoted by $n_{\beta}=0, s=1$;  $n_{\beta}=1, s=2$;  and  $n_{\beta}=2, s=3$ respectively  for the present $X(3)$ model and the  $X(5)$ model in ref.$^{27}$.}}
\begin{small}
\begin{center}
\begin{tabular}{ccccccccccc}
\hline 
&$\beta_{0}=2$&$\beta_{0}=2$&$\beta_{0}=3$&$\beta_{0}=3$&$\beta_{0}=4$&$\beta_{0}=4$&$\beta_{0}=15$&$\beta_{0}=15$\\
 \hline
$L$ &&&& &$n_{\beta}=0; s=1$ &&&\\
  
\hline
&$X(3)$&$X(5)$&$X(3)$&$X(5)$& $X(3)$ &$X(5)$&$X(3)$&$X(5)$\\
\hline
0&2.500 &2.031&2.803&  2.146&3.062&	2.250&4.905&3.077 \\
2&3.062 &2.250&3.291&  2.346&3.500&2.436&5.153&3.194 \\
4&3.986 &2.652&4.149&  2.726&4.304&2.797&5.682&3.445 \\
6&5.031 &3.136&5.153&  3.194&5.272&3.250&6.408&3.795 \\
8&6.123 &3.658&6.220&  3.704&6.315&3.750&7.265&4.212 \\
10&7.238&4.198&7.318&  4.237&7.397&4.276&8.205&4.671 \\

\hline
&&&& &$n_{\beta}=1;s=2$ &&&\\

\hline
0 &4.500&4.031&4.803& 4.146&5.062& 4.250&6.905& 5.077\\
2 &5.062&4.250&5.291& 4.346&5.500& 4.436&7.153& 5.194\\
4 &5.986&4.652&6.149& 4.726&6.304& 4.797&7.682& 5.445\\
6 &7.031&5.136&7.153& 5.194&7.272& 5.250&8.408& 5.795\\
8 &8.123&5.658&8.220& 5.704&8.315& 5.750&9.265& 6.211\\
10&9.238&6.198&9.318& 6.237&9.397& 6.276&10.205& 6.671\\

\hline
&&&& &$n_{\beta}=2; s=3$ &&&\\

\hline
0 &6.500&6.031&6.803&6.146&7.062&6.250&8.905& 7.077\\
2 &7.062&6.250&7.291& 6.346&7.500&6.436&9.153&7.194\\
4 &7.986&6.652&8.149& 6.726&8.304&6.797&9.682& 7.445\\
6 &9.031&7.136&9.153& 7.194&9.272&7.250&10.408& 7.795\\
8 &10.123&7.658&10.220& 7.704&10.315&7.750&11.265&8.211\\
10&11.238&8.198&11.318& 8.237&11.397&8.276&12.205& 8.671\\

\hline
\end{tabular}
\end{center}
\end{small}
\end{table}

\begin{figure}[hbtp]
\centering
\includegraphics[scale=0.4]{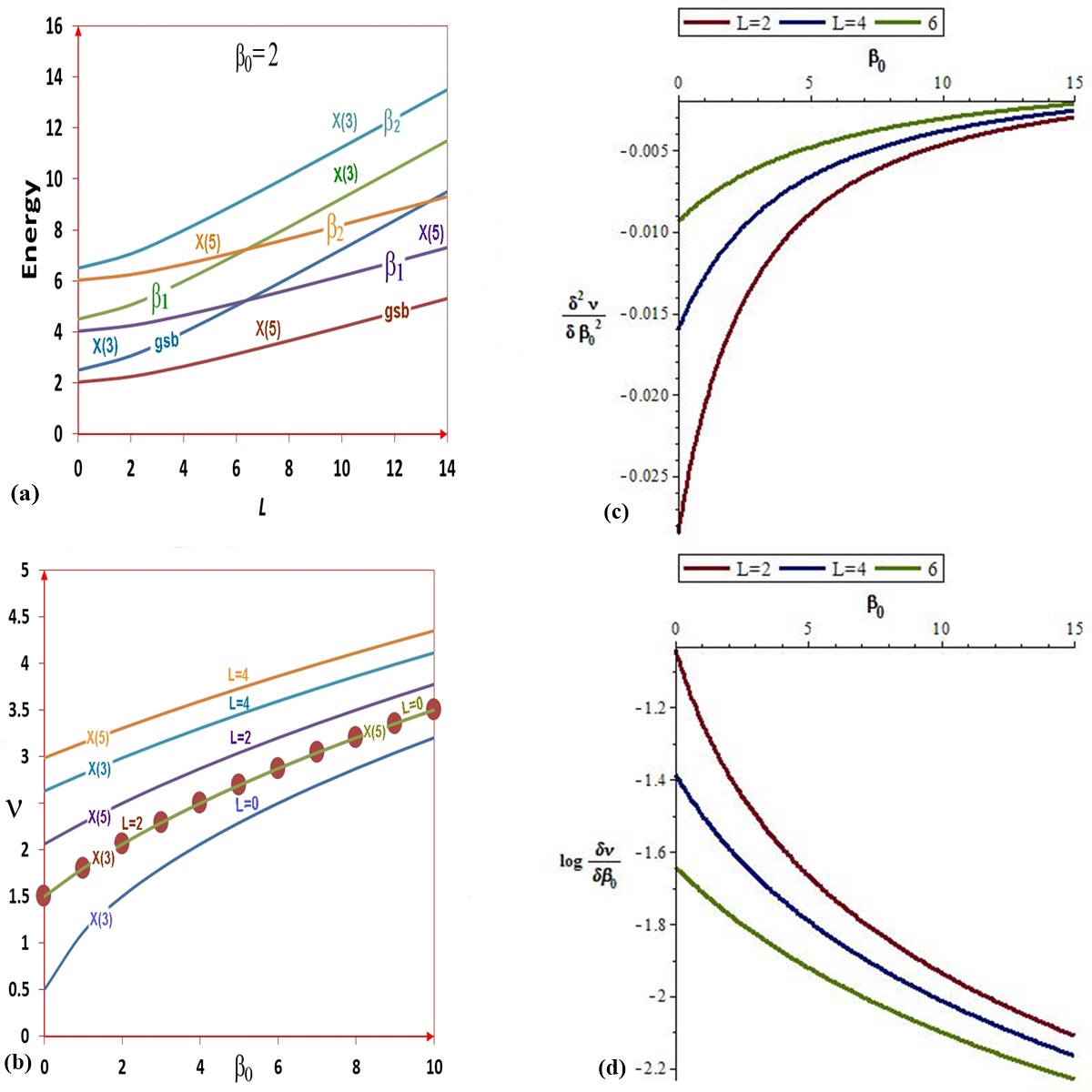}
\caption{{\small (a)  Comparison in the energy levels of the $X(3)$ and the $X(5)$ models in ref.$^{27}$  at $\beta_{0}=2$  from the $gsb$ up to the $\beta_{2}$-excited states. (b) The variation of the critical order, $\nu^{X(5)}$ in ref.$^{27}$ as a function of $\beta_{0}$, is compared with $\nu^{X(3)}$  model at different angular momentum: $L=0, 2$ and  $L=4$. (c) The plots show the procedure from which the value of $\beta_{0}$ that corresponds to the minimum energy for each level is obtained. (d) The plot shows the rate at which the energy level  changes with respect to $\beta_{0}$. This shows the non stationary property of $\beta_{0}$ in the description of critical point symmetries of isotopes.}}
\end{figure}

\begin{table}[hbtp]
\caption{{\small The ground state spectra ratios of the  present $X(3)$ model, at  different values of $\beta_{0}$,  are compared with the $X(5)$ model in ref.$^{27}$ in the first and the second sections of the table.  It can be seen that both $X(3)(\beta_{0}= \infty)$ and $X(5)(\beta_{0}= \infty)$, both with $4_{1,0}=3.296$ and  marked by the `$\dagger$' sign, approach $SU(3)$ whose  `Rotational' excitation signature$^{1}$, $4_{1,0}=3.333$. The third section compares the spectra ratios in the ground state and the $\beta$-excited states of the present model at different values of $\beta_{0}$  with the   $U(5)$ in ref.$^{29}$, the $X(5)$ in ref.$^{2}$ and the $SU(3)$ reported in ref.$^{30}$. `-' indicates the level where there is no data for comparison.}}
\begin{small}
\begin{center}
\begin{tabular}{ccccccccccccc}
\hline 	

$L_{s,n_{\beta}}$ &$\beta_{0}=0$&$\beta_{0}=0$&$\beta_{0}=2$&$\beta_{0}=2$&$\beta_{0}=4$&$\beta_{0}=4$&$\beta_{0}=\infty$&$\beta_{0}=\infty$&&	\\
 \hline
 &$X(3)$&$X(5)$&$X(3)$&$X(5)$&$X(3)$&$X(5)$&$X(3)$&$X(5)$ &&		\\
 \hline
  
$0_{1,0}$ &0.000&0.000&0.000&0.000&0.000&	0.000 &0.000&0.000&&\\
$2_{1,0}$ &1.000&1.000&1.000&  1.000&1.000&1.000 &1.000&1.000&&\\
$4_{1,0}$ &2.130&2.646&2.646&2.834&2.834&2.938&$^{\dagger}$3.296 &$^{\dagger}$3.296&&\\
$6_{1,0}$ &3.275&4.507&4.507&5.042&5.042&5.372&6.806&6.808&&\\
$8_{1,0}$ &4.424&6.453&6.453&7.421&7.421&8.508&11.413&11.423&&\\
$10_{1,0}$&5.576&8.438&8.438&9.887&9.887&11.881&16.991&17.013&&\\
$12_{1,0}$&6.728&10.445&10.445&12.404&12.404&15.686&23.409&23.450&&\\
$14_{1,0}$&7.882&12.465&12.465&14.951&14.951&19.740&30.544&30.611&&\\
 \hline
 &$\beta_{0}=1$&$\beta_{0}=1$&$\beta_{0}=3$&$\beta_{0}=3$&$\beta_{0}=5$&$\beta_{0}=5$&$\beta_{0}=15$&$\beta_{0}=15$&&	\\
 \hline

$0_{1,0}$ &0.000&0.000&0.000& 0.000&0.000&0.000&0.000&0.000&& \\
$2_{1,0}$ &1.000&1.000&1.000&1.000&1.000&1.000&1.000&1.000 &&\\
$4_{1,0}$ &2.476&2.756&2.756&2.893&2.893&2.946&3.128&3.148&&\\
$6_{1,0}$ &4.070&4.812&4.812&5.224&5.224&5.529&6.058&6.136&& \\
$8_{1,0}$ &5.706&6.995&6.995&7.767&7.767&8.638&9.508&9.690 &&\\
$10_{1,0}$&7.360&9.243&9.243&10.424&10.424&11.915&13.297&13.620&&\\
$12_{1,0}$&9.024&11.525&11.525&13.145&13.145&15.854&17.307&17.800&&\\
$14_{1,0}$&10.694&13.829&13.829&15.907&15.907&19.899&21.468&22.152&&\\
 \hline
& $\beta_{0}=0$&$\beta_{0}=1$&$\beta_{0}=2$&$\beta_{0}=3$&$\beta_{0}=4$&$\beta_{0}=15$&$\beta_{0}=\infty$&	$U(5)$&$X(5)$&$SU(3)$\\
 \hline
 
$0_{1,0}$  &0.000&0.000&0.000&0.000&0.000&0.000&0.000&0.000&0.000&0.000\\
$2_{1,0}$ &1.000&1.000&1.000&1.000&1.000&1.000&1.000&1.000&1.000&1.000\\
$4_{1,0}$ &2.130&2.476&2.646&2.756&2.834&3.128&$^{\dagger}$3.296&2.000&2.910&3.333\\
$6_{1,0}$  &3.275&4.070&4.507&4.812&5.042&6.058&6.806&3.000&5.450&7.000\\
$8_{1,0}$  &4.424&5.706&6.453&6.995&7.421&9.508&11.413&4.000&8.510&12.000\\
$10_{1,0}$  &5.576&7.360&8.438&9.243&9.887&13.297&16.991&5.000&12.700&18.333\\
$12_{1,0}$  &6.728&9.024&10.445&11.525&12.404&17.307&23.409&6.000&-&26.000\\
$14_{1,0}$  &7.882&10.694&12.465&13.829&14.951&21.468&30.544&7.000&-&35.000\\

$0_{2,1}$ &2.000&2.921&3.562&4.094&4.562&8.058&20.124&2.000& 5.670&-\\
$2_{2,1}$ &3.000&3.921&4.562&5.094&5.562&9.058&21.124&3.000&7.480&-\\
$4_{2,1}$ &4.130&5.397&6.208&6.850&7.395&11.187&23.420&4.000&10.720&-\\
$6_{2,1}$ &5.275&6.991&8.069&8.906&9.603&14.115&26.929&5.000&14.820&-\\
$8_{2,1}$ &6.424&8.626&10.014&11.090&11.982&17.567&31.537&6.000&-&-\\
$10_{2,1}$&7.576&10.281&11.999&13.337&14.449&21.356&37.116&7.000&-&-\\
$12_{2,1}$&8.728&11.945&14.007&15.619&16.965&25.366&43.534&-&-&-\\
$14_{2,1}$&9.882&13.615&16.027&17.923&19.513&29.526&50.668&-&-&-\\
  
$0_{3,2}$ &4.000&5.842&7.123&8.188&9.123&16.117&40.249&4.000&14.170&-\\
$2_{3,2}$ &5.000&6.842&8.123&9.188&10.123&17.117&41.249&5.000&16.780&-\\
$4_{3,2}$ &6.130&8.318&9.769&10.944&11.957&19.245&43.545&6.000&21.340&-\\
$6_{3,2}$ &7.275&9.912&11.630&13.000&14.165&22.174&47.054&7.000&-&-\\
$8_{3,2}$ &8.424&11.547&13.576&15.184&16.544&25.625&51.662&8.000&-&-\\
$10_{3,2}$&9.576&13.201&15.561&17.431&19.010&29.414&57.240&9.000&-&-\\
$12_{3,2}$&10.728&14.866&17.568&19.713&21.527&33.424&63.658&-&-&-\\
$14_{3,2}$&11.882&16.536&19.589&22.018&24.074&37.584&70.792&-&-&-\\
 \hline
 
\end{tabular}
\end{center}
\end{small}
\end{table}

\begin{figure}[hbtp]
\centering
\includegraphics[scale=0.26]{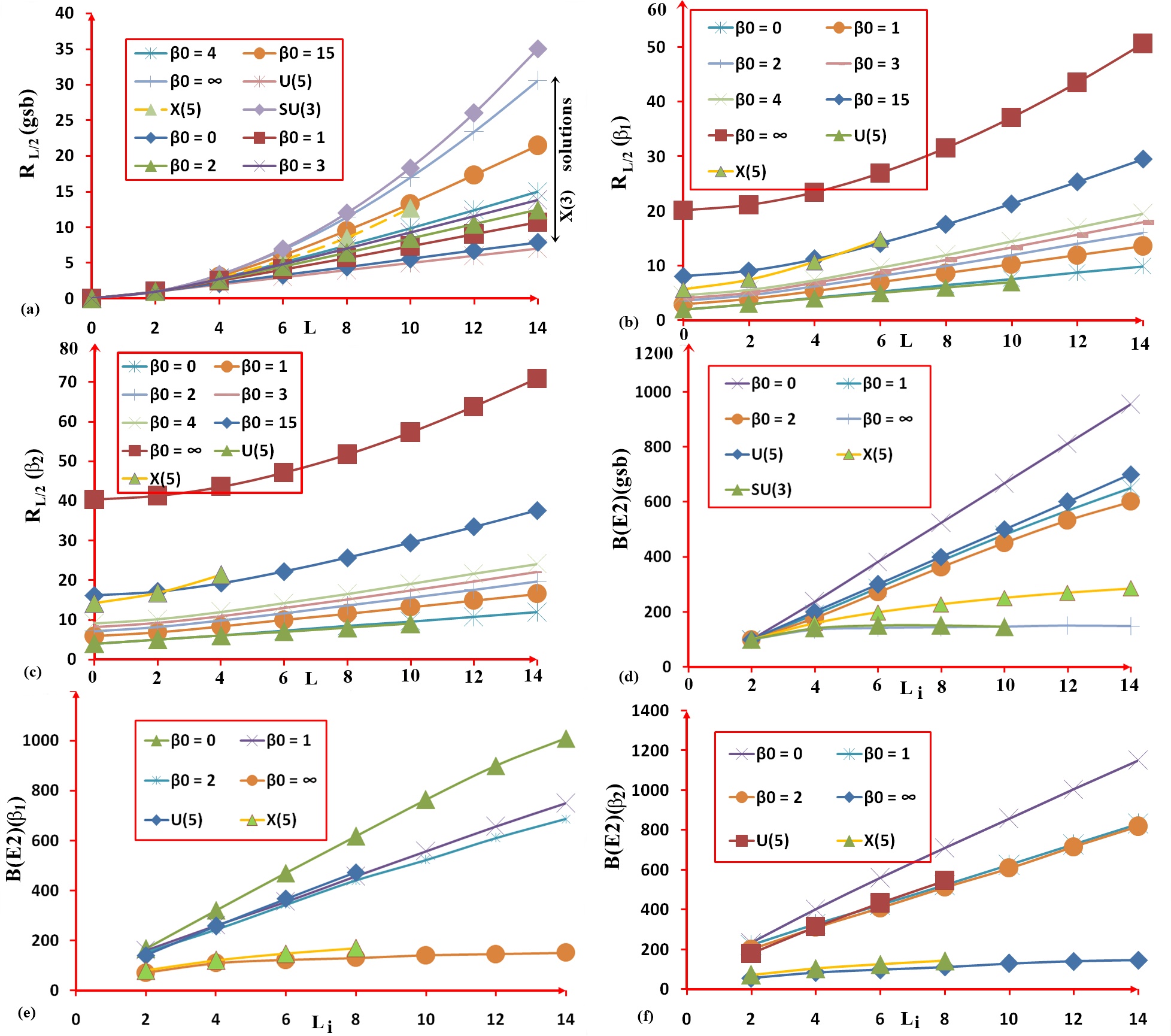}
\caption{The spectra ratios of the present $X(3)$ model plotted against the angular momentum for different values of $\beta_{0}$ are placed for comparison with the $U(5)$ in ref.$^{29}$, the $X(5)$ in ref.$^{29}$ and the $SU(3)$ reported in ref.$^{30}$: the comparison in the ground state and the  $\beta$-excited states are shown in (a), (b) and (c) respectively. (d) the  the ground state   (e) the $\beta_{1}$-excited states  and (f) the $\beta_{2}$-excited states: of the $B(E2)$   plotted against the initial angular momentum and calculated for different values of  $\beta_{0}$, are  respectively compared  with the $U(5)$ in ref.$^{29}$, the $X(5)$ in ref.$^{29}$ and the $SU(3)$ in ref.$^{30}$.}
\end{figure}

\begin{table}[hbtp]     
\caption{The distinct value of the $\beta_{0}$ that corresponds to each angular momentum obtained via $\beta_{0}$-optimization scheme$^{12,14,26,27,31}$ are labeled $\beta_{0,max}$. The calculated values of the $X(3)$ as a function of $\beta_{0,max}$,  labeled $X(3)$-var are compared with $X(3)$-IW models in ref.$^{8}$, $X(3)-\beta^{6}$ found in ref.$^{10}$ and $X(3)$-D in ref.$^{12}$ [Note: IW denotes infinite square well potential; D denotes Davidson potential; var denotes variational method; `-' indicates the level where there is no data for comparison.]}
\begin{small}
\begin{center}
\begin{tabular}{cccccccccccc}
 \hline	
$L_{s,n_{\beta}}$ &$\beta_{0,max}$&$X(3)$&$X(3)$&$X(3)$-D&$X(3)$&$L_{s,n_{\beta}}$ &$\beta_{0,max}$&$X(3)$&$X(3)$&$X(3)$-D&$X(3)$\\
 &&var&IW&var&$\beta^{6}$&&&var&IW&var&$\beta^{6}$\\
 \hline

$0_{1,0}$&$\beta_{0}$&0.000&0.000&0.000&0.000&$0_{3,2}$ &2.524&7.701&7.650&-&6.013\\
$2_{1,0}$&$\beta_{0}$&1.000&1.000&1.000&1.000&$2_{3,2}$ &4.497&10.553&10.560&-&7.925\\
$4_{1,0}$&0.844&2.440&2.440&2.570&2.343&$4_{3,2}$ &6.523&14.088&14.190&-&10.193\\
$6_{1,0}$&1.576&4.244&4.230&4.620&3.905&$6_{3,2}$ &8.567&18.172&18.220&-&12.606\\
$8_{1,0}$&2.033&6.383&6.350&7.130&5.654&$8_{3,2}$ &10.438&22.613&22.620&-&15.141\\
$10_{1,0}$&2.143&8.666&8.780&10.050&7.571&$10_{3,2}$&11.932&25.999&-&-&17.791\\
$12_{1,0}$&2.695&11.421&11.520&13.470&9.642&$12_{3,2}$&13.011&28.928&-&-&20.551\\
$14_{1,0}$&3.643&14.573&14.570&17.300&-&$6_{2,1}$ &5.213&10.327&10.290&-&7.968\\

$0_{2,1}$ &0.815&2.703&2.870&-&2.556&$8_{2,1}$ &6.098&13.493&13.570&-&10.141\\
$2_{2,1}$ &2.101&4.619&4.830&-&4.080&$10_{2,1}$&6.855&16.908&17.180&-&12.449\\
$4_{2,1}$ &3.729&7.255&7.370&-&5.941&$12_{2,1}$&8.106&21.009&21.140&-&14.884\\

\hline
\end{tabular}
\end{center}
\end{small}
\end{table}
 
\begin{table}[hbtp]     
\caption{{\small The first section of the table presents the $B(E2)$ transition rates of the present $X(3)$  model at $\beta_{0} = 0,1, 2$ and  $\beta_{0}=\infty$, normalized to the  $B(E2;2_{1,0}\rightarrow 0_{1,0})=100$ units are compared with the $U(5)$ in ref.$^{29}$, $X(5)$ in ref.$^{2}$ and $SU(3)$ reported in ref.$^{30}$. In the  second section of the table, the $B(E2)$-var obtained are compared with the  $B(E2)$ solutions obtained from the infinite square well potential, $X(3)$-IW, in ref$^{8}$. `-' indicates the level where there is no data for comparison.}}
\begin{footnotesize}
\begin{center}
\begin{tabular}{ccccccccccccccc}
\hline
	
$L_{s,n_{\beta}}^{(i)}$ &$L_{s,n_{\beta}}^{(f)}$&$\beta_{0}=0$&$\beta_{0}=1$&$\beta_{0}=2$&$\beta_{0}=\infty$&$U(5)$&$X(5)$&$SU(3)$ & & &&\\

 \hline

$2_{1,0}$&$0_{1,0}$ &100.000&100.000&100.000 &100.000&100.000&100.000&100.000&& &&\\
$4_{1,0}$&$2_{1,0}$ &207.941&190.935&178.005 &136.126&200.000&159.890&140.650&&  &&\\
$6_{1,0}$&$4_{1,0}$ &328.565&286.006&270.996 &141.873&300.000&198.220&150.540&&   &&\\
$8_{1,0}$&$6_{1,0}$ &472.664&384.599&369.090 &143.368&400.000&227.600&150.970&&   &&\\
$10_{1,0}$&$8_{1,0}$&589.862&486.036&469.991 &146.653&500.000&250.850&146.190&&  &&\\
$12_{1,0}$&$10_{1,0}$&693.722&587.658&559.744&152.134&600.000&269.730&-&&   &&\\
$14_{1,0}$&$12_{1,0}$&802.769&690.364&671.484&159.359&700.000&285.420&-&&   &&\\

$2_{2,1}$&$0_{2,1}$ &156.813&160.292&152.428&69.929&140.000&79.520&-&&  &&\\
$4_{2,1}$&$2_{2,1}$ &310.619&260.000&243.186&109.222&257.140&120.020&-&&  &&\\
$6_{2,1}$&$4_{2,1}$ &463.542&355.240&343.376&121.324&366.670&146.750&-&&   &&\\
$8_{2,1}$&$6_{2,1}$ &591.111&456.792&439.962&130.219&472.730&169.310&-&&    &&\\
$10_{2,1}$&$8_{2,1}$&718.669&557.317&542.129&139.999&-&-&-&&     &&\\
$12_{2,1}$&$10_{2,1}$&876.003&656.999&639.677&145.282&-&-&-&&    &&\\
$14_{2,1}$&$12_{2,1}$&1017.401&759.642&736.660&151.009&-&-&-&&    &&\\

$2_{3,2}$&$0_{3,2}$ &233.504&221.942&209.888&56.684&180.000&72.520&-&&   &&\\
$4_{3,2}$&$2_{3,2}$ &401.982&327.461&311.072&74.436&314.290&104.360&-&&   &&\\
$6_{3,2}$&$4_{3,2}$ &559.801&422.880&408.564&98.452&433.330&124.810&-&&    &&\\
$8_{3,2}$&$6_{3,2}$ &708.989&523.436&512.997&119.534&545.450&142.940&-&&    &&\\
$10_{3,2}$&$8_{3,2}$&858.095&624.555&609.096&129.234&-&-&-&&                &&\\
$12_{3,2}$&$10_{3,2}$&1003.933&727.909&715.990&140.274&-&-&-&&               &&\\
$14_{3,2}$&$12_{3,2}$&1151.239&832.003&819.115&158.278&-&-&-&&               &&\\

\hline
$L_{s,n_{\beta}}^{(i)}$ &$L_{s,n_{\beta}}^{(f)}$&$\beta_{0,max}^{(i)}$& $\beta_{0,max}^{(f)}$&$B(E2)$&$X(3)$&$L_{s,n_{\beta}}^{(i)}$ &$L_{s,n_{\beta}}^{(f)}$&$\beta_{0,max}^{(i)}$& $\beta_{0,max}^{(f)}$&$B(E2)$&$X(3)$\\
 &&&&-var&-IW& & &&&-var&-IW\\
\hline

$2_{1,0}$&$0_{1,0}$ &$\beta_{0}$& $\beta_{0}$&100.000&100.00&$10_{2,1}$&$8_{2,1}$&6.855& 6.098&242.026&242.40\\
$4_{1,0}$&$2_{1,0}$ &0.844& $\beta_{0}$&189.495&189.90&$12_{2,1}$&$10_{2,1}$&8.106& 6.855&268.543&265.10\\
$6_{1,0}$&$4_{1,0}$ &1.576& 0.844&250.995&248.90&$14_{2,1}$&$12_{2,1}$&9.441& 8.106&281.32&-\\
$8_{1,0}$&$6_{1,0}$ &2.033& 1.576&293.038&291.40&$2_{3,2}$&$0_{3,2}$ &4.497& 2.524&72.090&73.50\\
$10_{1,0}$&$8_{1,0}$&2.143& 2.033&324.599&323.80&$4_{3,2}$&$2_{3,2}$ &6.523& 4.497&118.990&120.50\\
$12_{1,0}$&$10_{1,0}$&2.695& 2.143&350.710&349.50&$6_{3,2}$&$4_{3,2}$ &8.567& 6.523&154.892&154.20\\
$14_{1,0}$&$12_{1,0}$&3.643& 2.695&371.992&370.70&$8_{3,2}$&$6_{3,2}$ &10.438& 8.567&183.019&181.20\\
$2_{2,1}$&$0_{2,1}$ &2.011& 0.815&78.922&80.60&$10_{3,2}$&$8_{3,2}$&11.932& 10.438&202.222&-\\
$4_{2,1}$&$2_{2,1}$ &3.729& 2.101&139.471&140.10&$12_{3,2}$&$10_{3,2}$&13.011& 11.932&218.753&-\\
$6_{2,1}$&$4_{2,1}$ &5.213& 3.729&181.730&182.40&$14_{3,2}$&$12_{3,2}$&14.629& 13.011&229.986&-\\
$8_{2,1}$&$6_{2,1}$ &6.098& 5.213&213.899&215.50&&&&&& &&\\

\hline
\end{tabular}
\end{center}
\end{footnotesize}
\end{table}

\begin{table}[hbtp]     
\caption{The numerical application of the spectra ratios of the present model is extended to twelve critical point  isotopes: $^{102}$Mo ref.$^{33}$, $^{104-108}$Ru  chain in ref.$^{34-36}$,  $^{120-126}$Xe chain in ref$^{37-40}$,  $^{148}$Nd in ref.$^{41}$, $^{184-188}$Pt chain in ref.$^{42-44}$ and also given to three $X(5)$ candidates: $^{150}$Nd in ref.$^{45}$, $^{154}$Gd in ref.$^{46}$ and $^{156}$Dy in ref.$^{47}$ with the  experimental data in the ground state, the first $\beta$-excited states and few data in the second $\beta$-excited states.  The values of the $\beta_{0}$ and the quality factor, $\sigma$, used during the fittings are recorded. `-' indicates the level where there is no data for comparison.}
\begin{footnotesize}

\begin{center}
\begin{tabular}{cccccccccccccc}
\hline\\ 
$L_{s,n_{\beta}}$ &$^{102}$Mo&$^{102}$Mo&$^{104}$Ru&$^{104}$Ru&$^{106}$Ru&$^{106}$Ru&$^{108}$Ru&$^{108}$Ru&$^{120}$Xe&$^{120}$Xe&$^{122}$Xe&$^{122}$Xe\\
&Exp&Theor&Exp&Theor&Exp&Theor&Exp&Theor&Exp&Theor&Exp&&\\
 \hline

$4_{1,0}$&2.510&2.566&2.480&2.468&2.660&2.662&2.750&2.623&2.470&2.522&2.500&2.571\\
$6_{1,0}$&4.480&4.483&4.350&4.299&4.800&4.805&5.120&5.102&4.330&4.263&4.430&4.500 \\
$8_{1,0}$&6.810&6.703&6.480&6.488&7.310& 7.199&8.020&7.999&6.510&6.361&6.690&6.759 \\
$10_{1,0}$&9.410&8.989&8.690&8.702&10.020&9.920&11.310&11.495&8.900&8.497&9.180& 9.216\\
$12_{1,0}$&-&11.498&-&11.878&-& 12.334&-&12.879&-&10.362&-&11.985\\
$14_{1,0}$&-&13.302&-&14.522&-& 15.073&-&15.591&-&12.640&-&14.759\\

$0_{2,1}$ &2.350&3.009&-&2.569&3.670&3.678&-&4.462&2.820&3.000&3.470&3.492\\
$2_{2,1}$ &3.860&4.301&4.230&4.233&-&5.127&-&5.902&3.950&4.203&4.510&4.608\\
$4_{2,1}$ &-&6.289&5.810&5.921&-&7.443&-&8.285&5.310&5.899&-&6.792\\
$6_{2,1}$ &-&8.691&-&8.009&-&10.001&-&11.099&-&7.958&-&9.172\\
$8_{2,1}$ &-&11.319&-&11.101&-&12.519&-&14.532&-&10.739&-& 11.829\\

$0_{3,2}$ &-&7.437&-&6.589&-&8.388&-&10.099&-&6.664&-&7.722\\
$2_{3,2}$ &-&9.000&-&8.202&-&10.200&-&11.811&-&8.007&-&9.402\\
$4_{3,2}$ &-&11.009&-&10.341&-&12.049&-&12.972&-&10.229&-&11.538\\
\hline

$\beta_{0}$&&1.464&&0.963&-&2.121&&1.830&&1.221&&1.494\\
$\sigma$&&0.569&&0.310&-&0.422&&0.276&&0.481&&0.295\\
\hline

&$^{124}$Xe&$^{124}$Xe&$^{126}$Xe&$^{126}$Xe&$^{148}$Nd&$^{148}$Nd&$^{184}$Pt&$^{184}$Pt&$^{186}$Pt&$^{186}$Pt&$^{188}$Pt&$^{188}$Pt\\
$L_{s,n_{\beta}}$&Exp&Theor&Exp&Theor&Exp&Theor&Exp&Theor&Exp&Theor&Exp&Theor\\
 \hline
$4_{1,0}$&2.480&2.498&2.420&2.463&2.490&2.500&2.670&2.721&2.560&2.567&2.530&2.890\\
$6_{1,0}$&4.370&4.418&4.210&4.291&4.240&4.187&4.900&5.001&4.580&4.399&4.460&4.503\\
$8_{1,0}$&6.580&6.596&6.270&6.325&6.150&6.007&7.550&7.679&7.010&6.768&6.710&6.666\\
$10_{1,0}$&8.960&8.726&8.640&8.382&8.190&7.986&10.470&10.585&9.700&8.863&9.180&8.969\\
$12_{1,0}$&-&11.998&-&10.479&-&9.769&-&12.807&-&11.999&-&11.378\\
$14_{1,0}$&-&15.079&-&12.681&-&11.246&-&15.367&-&14.875&-&14.242\\
$0_{2,1}$ &3.580&3.402&3.380&3.201&3.040&3.082&3.020&3.546&2.460&2.798&3.010 &3.209\\
$2_{2,1}$ &4.600&4.498&4.320&4.241&3.880&4.006&5.180&5.452&4.170&4.381&4.200 &4.397\\
$4_{2,1}$ &5.690&6.289&5.250&5.862&5.320&5.589&7.570&7.943&6.380&6.599&-&6.583\\
$6_{2,1}$ &-&8.564&-&7.828&7.120&7.411&11.040&11.157&8.360&8.581&-&8.603\\
$8_{2,1}$ &-&11.900&-&10.062&-&9.752&-&14.601&-&12.007&-&12.100\\
$0_{3,2}$ &-&7.334&-&6.759&-&6.249&-&10.122&-&7.389&-&7.603\\
$2_{3,2}$ &-&8.888&-&8.009&-&7.442&-&11.900&-&8.942&-&9.162\\
$4_{3,2}$ &-&10.022&-&9.287&-&9.051&-&12.998&-&10.208&-&10.441\\
\hline
$\beta_{0}$&&1.101&-&0.949&&1.111&&2.639&&1.469&&4.950\\
$\sigma$&&0.279&-&0.299&&0.347&&0.475&&0.600&&0.729\\
\hline

&$^{150}$Nd&$^{150}$Nd&$^{154}$Gd&$^{154}$Gd&$^{156}$Dy&$^{156}$Dy&&&&&&\\
$L_{s,n_{\beta}}$&Exp&Theor&Exp&Theor&Exp&Theor&&&&&&\\
 \hline
$4_{1,0}$&2.930& 2.501&3.010&2.483&        2.930&2.466&&&&&&\\
$6_{1,0}$&5.530& 4.696&5.830&4.392&        5.590&4.593&&&&&&\\
$8_{1,0}$&8.680& 7.576&9.300&7.375&        8.820& 7.410 &&&&&&\\
$10_{1,0}$&12.280&10.711&13.300&10.324&     12.520& 9.699&&&&&&\\

$0_{2,1}$ &5.190&4.958&5.530&5.521&        4.900&4.561 &&&&&&\\
$2_{2,1}$ &6.530&6.891&6.630&6.742&        6.010&6.268 &&&&&&\\
$4_{2,1}$ &8.740&9.010&8.510&8.839&        7.900&8.371&&&&&&\\
$6_{2,1}$ &11.830&12.004&11.100&11.008&    10.430&11.033&&&&&&\\
$8_{2,1}$ &-&13.857&-&13.062&              -&13.682&&&&&&\\

$0_{3,2}$ &(13.350)&10.633&9.600&10.202&   10.000& 9.863 &&&&&&\\
$2_{3,2}$ &-&13.148&11.52&11.899&             -&12.393 &&&&&&\\
$4_{3,2}$ &-&15.261&-&13.484&                 -&14.151&&&&&&\\
\hline
$\beta_{0}$&&4.284&-&0.956&&0.998&&&&&&\\
$\sigma$&&0.504&-&0.406&&0.408&&&&&&\\
\hline
\end{tabular}
\end{center}
\end{footnotesize}
\end{table}

\begin{table}[hbtp]     
\caption{The  ground states and the first $\beta$-excited states of the $B(E2)$  experimental data for some critical points symmetry isotopes:  $^{102}$Mo, $^{104}$Ru, $^{108}$Ru, $^{120}$Xe, $^{122}$Xe, $^{124}$Xe,  $^{148}$Nd and one X(5) candidate, $^{152}$Sm in ref.$^{48}$ are compared with the present $X(3)$ model. Except for $^{152}$Sm with the $\beta_{0}=1.052$ and the quality the factor $\sigma=0357$ which is fitted separately as done to  the isotopes in Table 6.,  the values of $\beta_{0}$ used for calculating the $B(E2)$ of other isotopes are taken from Table 6. `-' indicates the level where there is no data for comparison.}
\begin{small}
\begin{center}
\begin{tabular}{ccccccc}
\hline\\ 
Isotope &$4_{1,0}\rightarrow 2_{1,0}$&$6_{1,0}\rightarrow 4_{1,0}$&$8_{1,0}\rightarrow 6_{1,0}$&$10_{1,0}\longrightarrow 8_{1,0}$&$0_{2,1}\longrightarrow 2_{1,0}$&$2_{2,1}\longrightarrow 2_{1,0}$\\

 \hline
$^{102}$Mo&120.000&-&-&-&-&-\\
&178.323&259.875&326.632&371.284&222.228&39.736\\
$^{104}$Ru&143.000&-&-&-&-&-\\
&178.899&263.544&335.734&400.000&218.184&16.499\\

$^{108}$Ru&165.000&-&-&-&-&-\\
&170.006&223.246&270.642&309.663&200.453&40.235\\
$^{120}$Xe&116.000&117.000&96.000&91.000&-&-\\
&185.634&269.643&332.754&401.523&213.432&32.333\\
$^{122}$Xe&146.000&141.000&103.000&154.000&-&-\\
&180.564&244.444&300.634&356.544&230.531&41.483\\
$^{124}$Xe&117.000&152.000&114.000&36.000&-&-\\
&146.422&255.874&319.581&383.451&225.632&28.991\\

$^{148}$Nd&162.000&176.000&169.000&-&54.000&25.000\\
&191.045&265.354&349.274&442.632&13.573&24.745\\

$^{152}$Sm&144.000&166.000&202.000&217.000&23.000&4.000\\
&159.853&199.576&238.618&306.268&113.147&28.354\\

$^{184}$Pt&165.000&178.000&213.000&244.000&-&-\\
&167.521&217.007&271.326&284.532&188.809&28.611\\

\hline

\end{tabular}
\end{center}
\end{small}
\end{table}
 
\begin{figure}[hbtp]
\centering
\includegraphics[scale=0.21]{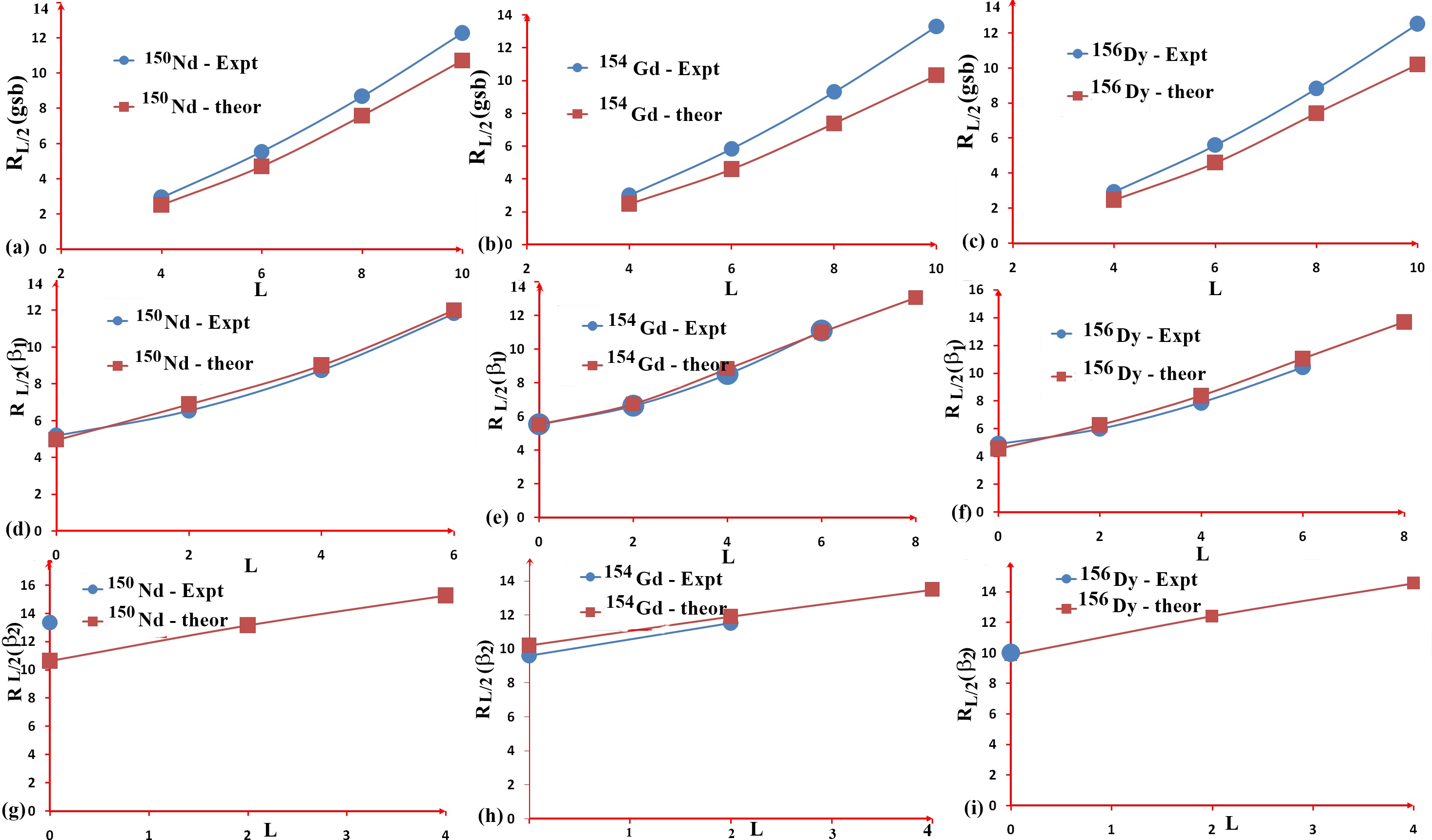}
\caption{The visual comparison between the experimental data of the $X(5)$ candidates isotopes: $^{150}$Nd in ref$^{45}$, $^{154}$Gd in ref$^{46}$, $^{156}$Dy in ref$^{47}$ and the spectra ratios of present $X(3)$ theoretical model  in the ground state, the $\beta_{1}$-excited states and the $\beta_{2}$-excited states.}
\end{figure}

\end{document}